\documentclass[a4paper,10pt,conference]{IEEEtran}
\usepackage{cite}
\usepackage{times}
\usepackage{dsfont}
\usepackage{graphicx}
\usepackage{subfigure}
\usepackage{color}
\usepackage{ifpdf}
\usepackage{epsfig}
\usepackage{latexsym}
\usepackage{amsfonts}
\usepackage{amssymb}
\usepackage{paralist}
\usepackage{comment}
\usepackage{xspace}
\usepackage{mathrsfs}
\usepackage{amssymb}
\usepackage{slashbox}
\usepackage{setspace}
\usepackage{color}
\usepackage[small]{caption2}

\usepackage{algorithmic}
\usepackage[ruled]{algorithm}
\usepackage{algorithmic}

\usepackage{amssymb}
\usepackage{url,epsfig,array}
\usepackage{leftidx}
\usepackage{amsmath}

\def\iid{\textit{i.i.d.}\xspace}

\newtheorem{theorem}{Theorem}

\newtheorem{lemma}[theorem]{Lemma}

\ifCLASSINFOpdf
\else
\fi
\begin{document}
%
\title{Queueing Analysis for Preemptive Transmission in Underlay Cognitive Radio Networks}
\author{\IEEEauthorblockN{Long Chen$^{\dag\ast}$ \   Liusheng Huang$^{\dag\ast}$  \ Hongli Xu$^{\dag\ast}$ \ Jie Hu$^{\dag\ast}$   }
\IEEEauthorblockA{Email:
lonchen@mail.ustc.edu.cn, lshuang@ustc.edu.cn, xuhongli@ustc.edu.cn, hujie826@mail.ustc.edu.cn \\
$^{\dag}$ School of Computer Science and Technology, University of
Science and Technology of China, Hefei, Anhui, P.R.China \\
$^{\ast}$ Suzhou Institute for Advanced Study, University of Science and Technology of China, Suzhou, Jiangsu, P.R.China
}}


%


\maketitle

\begin{abstract}\label{sec:abstract}
In many cognitive radio applications, there are multiple types of message queues. Existing queueing analysis works in underlay CR networks failed to discuss packets heterogeneity. Therefore high priority packets with impatient waiting time that have preemptive transmission opportunities over low class are investigated. We model the system behavior as a M/M/1+GI queue which is represented by a two dimensional state transition graph. The reneging probability of high priority packets and the average waiting time in two-class priority queues is analyzed. Simulation results demonstrate that the average waiting time of high priority packets decreases with the growing interference power threshold and the average waiting time of the low priority packet is proportional to the arrival rate of the high priority packet. This work may lay the foundation to design efficient MAC protocols and optimize long term system performance by carefully choosing system parameters.
\end{abstract}


%

%
\IEEEpeerreviewmaketitle

\section{Introduction}\label{sec:introduction}

Spectrum resources have rapidly become scarcity in recent years with the explosive growing number of wireless communication devices. However, some spectrums have not been fully utilized due to the exclusive spectrum usage in dedicated application scenarios \cite{huang2010beyond}. Cognitive Radio (CR) \cite{mitola2000cognitive} is a promising technology to solve the spectrum under-utilization problem and to mitigate the spectrum scarcity. It has been paid much attention ever since the year 2000. With the capability to sense, detect and access the frequency bands that are not being occupied currently, CR technology allows secondary users (SUs) (or unlicensed users) to exploit those spectrum bands which are unused by primary users (PUs) (or licensed users) in an opportunistic manner \cite{song2012distributed}. SUs grouped together can form both infrastructure based networks \cite{ozger2013event} and ad-hoc networks such as cognitive vehicular networks \cite{han2014throughput}.

CR networks (CRNs) can be classified into three main categories by the differentiation of spectrum utilization methodology. Namely, they are overlay approach, interweave approach and underlay approach \cite{srinivasa2007cognitive}. For the overlay approach, SUs utilize the same spectrum of PUs but only use a portion of power for secondary transmissions and the remainder power to assist PUs' transmissions. This is usually difficult to be implemented due to the sophisticated coding and power splitting methods. For the interweave approach, SUs can only utilize the bandwidth which is not currently being occupied by PUs and when PUs come back, SUs should vacate the channel immediately. This approach is used by users in an opportunistically manner and is not suitable for time critical communications. Finally, for the underlay approach, SUs are authorized to use the same spectrum occupied by PUs provided the interference power to PUs are within a threshold that PUs could tolerate. Underlay approach is used in numerous situations. For example, by deploying femto-cells underlying macro-cells, it is beneficial for enhancing the coverage of indoor communications as well as increasing system capacity \cite{cheng2011exploiting}. In cognitive vehicular networks, vehicles act as SUs \cite{ghandour2013improving} can  concurrently communicate with PUs along the roadside such that the interference powers to PUs are controlled. Therefore, in this paper we mainly focus on the underlay approach.

In CR networks, queueing based model can be used for cognitive system engineering \cite{rashid2009opportunistic} such as spectrum scheduling, admission controller design and so on. Only a few papers have addressed the queueing behavior in underlay CRNs, for instance, transmission delay, packet blocked probability and so on \cite{chu2013performance}\cite{tran2011queuing}\cite{suliman2009queueing}\cite{khabazian2011modeling}. The authors in \cite{chu2013performance} analyzed the performance of the CRN such as average packet transmission time, system throughput, average waiting time, average queue length etc. However, they assumed that all packets are homogeneous. In \cite{tran2011queuing}, M/G/1 queueing model was used to analyze the system performance which was similar as \cite{chu2013performance}. Both \cite{chu2013performance} and \cite{tran2011queuing} assumed the time-out waiting time was fixed, which couldn't reflect the randomness of impatience time. A M/D/1 queueing model was employed by \cite{suliman2009queueing} to analyze the performance of both PUs' and SUs' packets in an overlay CRN model, not in underlay CRNs. Meanwhile, the failure to exploit the packets' heterogeneity limits its applicability. Simulation results of \cite{suliman2009queueing} showed that the average waiting time of PUs grew with the number of PUs. Cooperative communication was adopted by \cite{khabazian2011modeling} and the queueing characteristics were illustrated in the overlay CRNs. In most recent work, \cite{kam2013multicast} studied the stability of transmission throughput in cooperative CRNs with multicast. What makes this paper distinct from \cite{chang2013spectrum} is that in \cite{chang2013spectrum} PUs possessed preemptive priority over SUs while in this paper, preemptive priority is owned by high priority packets over low priority packets in the SU network. Almost all the preceding literatures fail to take into account the heterogeneity priority of packets in underlay CRNs, which are not suitable for scenarios when there are heterogeneous packets in the SU network.

In reality, considering a cognitive vehicular network where both time critical and periodic messages coexist, safety related messages are much more urgent than non-safety related messages. Hence safety related messages should be granted with higher priority and be transmitted first while recreational or conventional messages should be processed afterwards. On the battle field \cite{wang2013belief}, when soldiers act as SUs moving among surveillance sensors, the messages sent by soldiers should be immediately handled prior to messages sent from static sensors. Since existing methods are not applicable to the above scenarios, we attempt to fill the gap between queueing analysis and heterogeneous priority packets' transmissions in the underlay CR network.

The main contributions of this paper are summarized as follows. We model the network as a M/M/1+GI queueing system with two-class priority queues and generally independently distributed impatient waiting time. Packets in the high priority queue have preemptive priority to be transmitted by the cognitive transmitting node. While low priority packets are permitted to transmit when the high priority queue is empty. Then we employ a two dimensional state transition graph to imitate the system queueing behavior. By solving the balanced equations of state transition graph for the two-class priority queues, we analyze the queueing performances such as average queueing delay, reneging probability, and system idle probability on the two queues through simulations. This work may lay the foundation of future cognitive communication system designs. To the best of the authors' knowledge, it is the first time to study the queueing characteristics of an underlay CRN with heterogenous priority transmission packets.

The remainder of this paper is organized as follows. In Section \ref{sec:systemmodel}, system model is presented including queueing, channel and impatience sub models . To model the system behavior, Section \ref{sec:modelingsystembehavior} illustrates a two dimensional state transition graph to imitate the stable state of the system. In Section \ref{sec:queueinganalysis}, the queueing characteristics of the two class high priority and low priority queues are presented. Simulation results are shown in Section \ref{sec:simulation} and we conclude this paper in Section \ref{sec:conclusion}.

\section{System Model}\label{sec:systemmodel}
\begin{figure}[h]\centering
\includegraphics[width=2.5in]{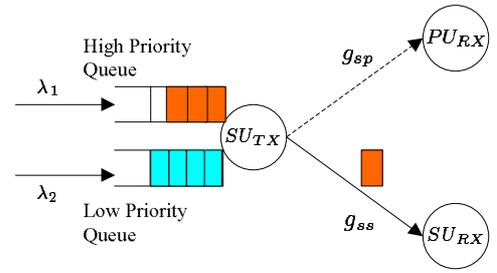}
\caption{System model for two-class queue underlay CRN}\label{Fig1}
\end{figure}

In this section, we present the system model as shown in Fig. \ref{Fig1}. There is one SU transmission node $SU_{TX}$ and one SU receive node $SU_{RX}$. $PU_{TX}$ is primary transmission node which is omitted in the figure and $PU_{RX}$ is the primary receive node. Two-class priority queues are deployed in the system. One is a high priority (or class-1) queue the other is low priority (or class-2) queue. Without loss of generality, we assume the capacity of the two queues are infinite. Also, we will give numerical analysis when the queues are finite due to storage limitations. In this model, the high priority queue has preemptive priority over the low priority queue. That is, whenever there are class-1 packets in the system, class-2 packets cannot be served. When one class-1 packet arrives and meanwhile one class-2 packet is being transmitted by $SU_{TX}$, the class-2 packet will immediately cease its transmission and come back to the head of class-2 queue. Then the transmission node $SU_{TX}$ will serve the coming class-1 packet. The previous assumption is meaningful. For example, in a cognitive vehicular network \cite{ghandour2013improving} when collision warning information \cite{garcia2012stochastic} are concurrently transmitted with periodic location related messages \cite{yu2007self}\cite{guo2013r}, the warning messages should have preemptive priority to be spread out over the location based messages. In this way, roadway safety could be improved.

\subsection{Queueing Model}
In the system, the arrival process of class-1 packets follows Poisson distribution with mean value $\lambda_1^{-1}$ and the arrival process of class-2 packets follows Poisson distribution with parameter $\lambda_2$. Both of the two kinds packets' arrival processes are independent identically distributed (\iid). The high priority packets have impatient time $t_{out}$ which follows general distribution \cite{ross2014introduction} such as Poisson distribution, Bernoulli distribution with parameter $\gamma$. If the class-1 packet fails to be transmitted by $SU_{TX}$ node within waiting time $t_{out}$, then it will renege from the current queue and search for other available channels to transmit. Since there have been numerous literatures concerning on spectrum handoffs in CRNs, how to perform spectrum handoff is out of the scope of this paper. The system presented in this paper can be viewed as one part of a very large cognitive radio network.

Let $\mu_i,i\in\{1,2\}$ denote the service rate of $SU_{TX}$, which follows exponential distribution. Among each of the two class queues, FCFS queueing rule is adopted. The $SU_{TX}$ in the system is the only transmission node. Therefore the system shown in Fig. \ref{Fig1} is a two-class M/M/1+GI queueing model, where GI is the time out period which follows generally independent distribution. According to Pollaczek-Khinchin (P-K) formula \cite{klennrock1975queueing}, the expectation of average waiting time $E[W_1]$ for class-1 packets in the system can be expressed as:
\begin{equation}\label{eq1}
 E[W_1]=E[T]+E[T_{q_1}]
\end{equation}where $W_1$ is the average waiting time for high priority packets in class-1 queue, $T$ is the average transmission time for high priority packets and $T_{q_1}$ is the average queueing delay in class-1 queue. To ensure low priority packets' QoS when being served by the $SU_{TX}$, the expectation of average waiting time in class-2 queue $E[T_{q_2}]$ should not exceed an upper bound $\epsilon$, hence:
\begin{equation}\label{eq2}
 E[T_{q_2}]< \epsilon
\end{equation}

\vspace{-0.5em}
\subsection{Channel Model}
To ensure the successful transmissions between PU nodes $PU_{TX}$ and $PU_{RX}$ when performing concurrent transmissions from $SU_{TX}$ to $SU_{RX}$ in the underlay CRN, the transmission power of $SU_{TX}$ should be limited so that the interference power received at primary receive node $PU_{RX}$ will not exceed its maximum tolerable interference power threshold $Q$. In this system, we assume $PU_{TX}$ is far away from SUs so that the interference caused by $PU_{TX}$ node can be neglected. For simplicity we assume transmission rates are equal for the two-class priority packets. In order to ensure the stability of transmission, the arrival rate of packets should not exceed the service rate \cite{gong2006queue}. Hence,
\begin{equation}\label{eq3}
\lambda_1+\lambda_2<\mu, (\mu_1=\mu_2=\mu)
\end{equation}

Next, we derive the average transmission time of $SU_{TX}$. For simplicity, we assume all packets are of the same length, denoted by $S$. Let $B$ stand for system bandwidth, $\gamma_{r_x}$ denote the signal to noise ratio ($SNR$) at the cognitive receive node $SU_{RX}$. Since the transmission time $T$ is inversely proportional to transmission rate, thus \cite{khan2012delay}:
\begin{equation}\label{eq4}
T={\frac{S}{B\log_{2}(1+\gamma_{r_x})}}= {\frac{\bar{B}}{\log_e(1+\gamma_{r_x})}}
\end{equation}where $\gamma_{r_x}=\frac{g_{ss}P_s}{N_0}$, $N_0$ is the variance of additive white Gaussian noise (AWGN) with zero mean and $\bar{B}$ is bandwidth-normalized entropy. In underlay cognitive radio scheme on one hand the $SU_{TX}$ requires a higher power to gain a higher transmission rate; on the other hand, if the transmission power is too large, then interference to the primary receive node $PU_{RX}$ will be high. Therefore, the transmission power $P_s$ of $SU_{TX}$ should be within the range that $PU_{RX}$ could tolerate, hence:
\begin{equation}\label{eq5}
P_s \le \frac{Q}{g_{sp}}
\end{equation}where $g_{sp}$ is the channel power gain between $SU_{TX}$ and $PU_{RX}$. Let $g_{ss}$ denote the channel power gain between $SU_{TX}$ and $SU_{RX}$ and by substituting the maximum interference power $Q$ of (\ref{eq5}) into the expression of $\gamma_{r_x}$, then (\ref{eq4}) can be rewritten as \cite{tran2013cognitive}:
\begin{equation}\label{eq6}
T=\frac{\bar{B}}{\log_e {(1+ {\frac{g_{ss}Q}{g_{sp}N_0} })}}
\end{equation}

According to \cite{tran2011queuing}, the probability density function (PDF) of transmission time at $SU_{TX}$ is:
\begin{equation}\label{eq7}
f_{T}(t)={\frac{{\bar{B}}{Q}}{N_{0} t^{2}}} { \frac{e^{\bar{B}/t}}{(Q/{N_0} -1 + e^{{\bar{B}}/t} })^2}
\end{equation}

\subsection{Impatience Model}
In this subsection, we introduce the impatience model for high priority packets in class-1 queue. When the high priority packet is not served within the impatient time $t_{out}$, it will leave its current queue and search for the remained channels for transmission. If $SU_{RX}$ receives a packet before $t_{out}$, the $SU_{RX}$ then sends an ACK message to the $SU_{TX}$. If $SU_{TX}$ does not receive the ACK message, it waits until the $t_{out}$ is met. The details of this scheme are illustrated in Algorithm \ref{alg:class1queue}.
{\small
\begin{algorithm}[h]
\caption{Impatience behavior of the high priority packet}\label{alg:class1queue}
\begin{algorithmic}[1]
\STATE \textbf{Process high priority packet in class-1 queue}
\WHILE {Class-1 queue is not empty}
 \FOR{each high priority packet}
   \IF{($waiting time < t_{out}$)}
     \IF{$SU_{TX}$ is idle}
        \STATE {Be served by $SU_{TX}$ immediately;}
      \ELSE
          \IF{$SU_{TX}$ is sending class-1 packet}
             \STATE {wait \& waiting time increases then go to line 4;}
           \ELSE
              \STATE {Replace the serving low priority packet and to be served by $SU_{TX}$ immediately;}
           \ENDIF
      \ENDIF
    \ELSE
      \STATE {Reneging;}
    \ENDIF
 \ENDFOR
\ENDWHILE
\end{algorithmic}
\end{algorithm}
}

The behavior of cognitive radio users $SU_{TX}$ and $SU_{RX}$ is explicitly depicted in Algorithm \ref{alg:sutxsurx}.
{\small
\begin{algorithm}[h]
\caption{Transmission behavior of cognitive radio users $SU_{TX}$ and $SU_{RX}$}\label{alg:sutxsurx}
\begin{algorithmic}[1]
\STATE \textbf{Behavior of $SU_{TX}$}
\WHILE {the service cache is not empty}
    \IF{(the packet is class-1 packet)}
       \STATE {Send the packet to cognitive receive node $SU_{RX}$;}
    \ENDIF
     \IF{received the ACK from $SU_{RX}$ within $t_{out}$}
        \STATE {process the next packet, go to line 2;}
     \ENDIF
     \STATE {notify the packet for reneging;}
     \IF{the packet is class-2 packet}
      \STATE {Send the packet to cognitive receive node $SU_{RX}$;}
      \ENDIF
\ENDWHILE
\STATE \textbf{Behavior of $SU_{RX}$}
    \IF{received the packet sent from $SU_{TX}$}
       \STATE {Send one ACK message to $SU_{TX}$;}
    \ENDIF
\end{algorithmic}
\end{algorithm}
}

Let $P_{out}$ denote the probability that the high priority packets are unsuccessfully sent \cite{tran2011queuing}\cite{khan2012delay}\cite{tran2012delay}. That is,
\begin{equation}\label{eq8}
P_{out}=Pr\{T>t_{out}\}=1- \frac{Q}{{N_0} ({e^{\frac{\bar{B}}{t_{out}}}}+Q/N_0 -1)}
\end{equation}
The expectation of average transmission time of high priority packets is:
\begin{equation}\label{eq9}
\begin{split}
E[T]=&\frac {Q}{N_{0}} [\int_{{e^{\frac{\bar{B}}{t_{out}}}}}^{\infty} \frac{\bar{B}}{(Q/N_0-1+u)^2 \log_{e}u} du+\\
&t_{out} (\frac{N_0}{Q}-\frac{1}{1-e^{\frac{\bar{B}}{t_{out}}}})]
\end{split}
\end{equation}

\section{Modeling the System Behavior}\label{sec:modelingsystembehavior}
\begin{figure}[h]\centering
\includegraphics[width=3.0in]{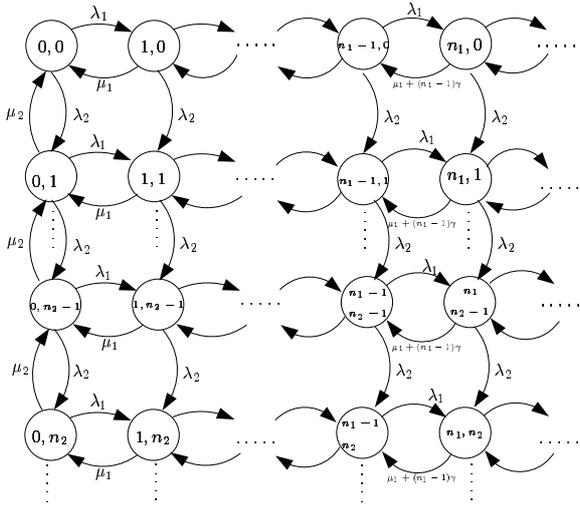}
\caption{State Transition Graph of the System}\label{Fig2}
\end{figure}

In this section, to derive the expressions of $E[T_{q_2}]$ and $E[T_{q_1}]$, the queueing behavior of the system will be analyzed. According to queueing theory, the behavior of the system can be modeled as a two dimensional state transition graph. The graph is shown in Fig. \ref{Fig2}. Let $n_1(t)$ and $n_2(t)$ be the number of class-1 and class-2 packets at time $t$ in the system accordingly. Considering the bivariate process $\{n_1(t),n_2(t),t\ge 0\}$ in the state space $S=\{(i,j):i,j=0,1,2,\cdots\}$, the stable probability of the system can be defined as: $P_{ij}=Pr\{$in steady-state there are $i$ class-1 packets and $j$ class-2 packets in the system$\}$. Before formulating the system behavior, we propose some lemmas which will be used during the formulation procedure.
\begin{lemma}\label{lemma:1}
There is a stable state distribution of the system.
\begin{IEEEproof}
The lemma can obviously be established. In reality, the storage of cognitive transmission node is finite, thus the queue length is finite. Therefore, the state space is also finite. In this way, we can determine that there is a stable state distribution of the system. Another way of proof based on literature \cite{gong2006queue} can be summarized as below. In \cite{gong2006queue}, the state transition graph is divided by the cut-off length $L$. When $L=0$, the sate transition graph just reduced to this state transition graph shown in Fig. \ref{Fig2}. Because there is a stable state distribution of \cite{gong2006queue}, hence there must be a stable distribution of the proposed system in this paper.
\end{IEEEproof}
\end{lemma}
\begin{lemma}\label{lemma:2}
When the queue length of high priority packet is determined, which is denoted by $n$, the reneging probability of packets in the high priority queue $P_{reneging}^{n}$ can be expressed as:
\begin{equation}\label{eq10}
P_{reneging}^{n}= \begin {cases}
  0, & \text{n=0} \\
  (n-1)\gamma, & \text{otherwise}
\end {cases}
\end{equation}
\begin{IEEEproof}
As shown in Fig. \ref{Fig2}, when state $(n_1,n_2)$ transforms to state $(n_1-1,n_2)$, according to Lemma \ref{lemma:2} the queue length of class-1 queue turns to $n_1-1$ from $n_1$, then $P_{reneging}^{n_1}=(n_1-1)\gamma$. This is true because when there are class-1 packets in the system, low priority packets cannot be served until the high priority queue is empty. Therefore, the reneging probabilities of class-1 packets are not in connection with the low priority queue. When the system is stable, there are $n_1$ class-1 packets in the system at state $(n_1,n_2)$. Hence, when it comes to state $(n_1-1,n_2)$, each of the $n_1-1$ waiting packets may be impatient. Since the impatient time of all those packets are \iid hence, the packet that possesses the lowest $t_{out}$ value will renege from the current high priority queue. It is easy to see that by mathematical induction, the lemma is proofed.
\end{IEEEproof}
\end{lemma}

According to Lemma \ref{lemma:1} and Lemma \ref{lemma:2}, the balance equations to the two dimensional state transition graph shown in Fig. \ref{Fig2} can be:
\begin{equation}\label{eq11}
\begin{split}
&i=0,j=0,   \quad \quad \quad \\
 (\lambda_1+ & \lambda_2)  P_{00} = \mu_1 P_{10}+\mu_2 P_{01};
\end{split}
\end{equation}
\begin{equation}\label{eq12}
\begin{split}
 &i=0,j>0, \quad \quad  \\
P_{0j}(\lambda_1+\lambda_2 & +\mu_2)=  \\
& \lambda_2 P_{0j-1}+\mu_2 P_{0j+1}+\mu_1 P_{i+1j};
\end{split}
\end{equation}
\begin{equation}\label{eq13}
\begin{split}
& i>0, j = 0,   \quad \quad \\
P_{i0}(\lambda_2+\lambda_1 & +\mu_1+(i-1)\gamma)= \\
& \lambda_1 P_{i-10}+(\mu_1+(i-1)\gamma) P_{i+10};
\end{split}
\end{equation}
\begin{equation}\label{eq14}
\begin{split}
&\quad \quad \quad \quad i>0, j>0,   \quad \quad \quad \\
P_{ij}(\mu_1+ & (i-1)\gamma+\lambda_1+\lambda_2)=  \\
& \lambda_2 P_{ij-1}+ {\lambda_1 P_{i-1j}} +{ (\mu_1+(i-1)\gamma)P_{ij+1} }
\end{split}
\end{equation}

It can be seen from (\ref{eq11})(\ref{eq12})(\ref{eq13})(\ref{eq14}), it is hard to solve the state balance equations using iterative method directly. Meanwhile, it is also hard to solve them using generating function method. We will give approximate solutions to the above equations in the next section.

\section{Queueing Analysis}\label{sec:queueinganalysis}
In this section, we will derive the properties of the two-class priority queueing system. The reneging probability, the expectation of the average waiting time in the high priority queue and the expectation of the average waiting time in the low priority queue are calculated.

\subsection{High Priority Queue Analysis}
\begin{figure}[h]\centering
\includegraphics[width=3.0in]{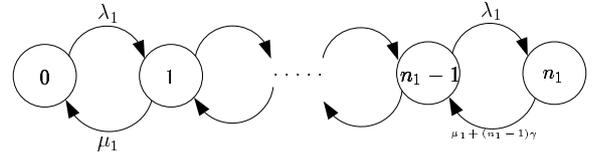}
\caption{State Transition Graph of High Priority Queue}\label{Fig3}
\end{figure}

Since class-1 packets are always first to be served by $SU_{TX}$ node, thus in case there is one packet in the high priority queue, it will be transmitted by the $SU_{TX}$. This transmission occurs without the influence of class-2 queue. Therefore, for class-1 packets the horizonal lines of the two dimensional graph make sense. Suppose the maximum capacity of class-1 queue is $n_1$, thus the M/M/1+GI queueing model can be reduced to a M/M/1/$n_1$ queueing model for the high priority packet queue. The reduced state transition graph for high priority queue is depicted in Fig. \ref{Fig3}.

The balance equation of Fig. \ref{Fig3} is:
\begin{equation}\label{eq15}
\lambda_1 P_{n-1} = [\mu_1 + (n-1)\gamma]P_{n}
\end{equation}which can be rewritten as:
\begin{equation}\label{eq16}
P_{n}=\frac{\lambda_1 P_{n-1}}{\mu_1 + (n-1) \gamma}, 1\le n \le n_1
\end{equation}

It can be obtained from (\ref{eq16}) through iterations with $\sum_{n=0}^{n_{1}}P_{n} =1
$:
\begin{equation}\label{eq17}
P_{0}=\frac {1}{1+\sum_{n=1}^{n_1} \prod_{i=1}^{n} \frac{\lambda_1}{\mu_1+(i-1)\gamma}}
\end{equation} where $P_0$ is the empty probability of the high priority queue and $P_n$ can be represented by $P_0$ as:
\begin{equation}\label{eq18}
P_{n}= \frac {{\lambda_1}^{n}} {\prod_{i=1}^{n}[\mu_1+(i-1)\gamma]} P_{0},1\le n \le n_1
\end{equation} Hence the average queue length $L_{q_1}$ of class-1 queue is:
\begin{equation}\label{eq19}
L_{q_1}=\sum_{n=0}^{n_1}n\cdot P_{n}
\end{equation}

According to Little's Law \cite{klennrock1975queueing},
\begin{equation}\label{eq20}
\lambda_1 E[T_{q_1}]=L_{q_1}
\end{equation}By substituting equations (\ref{eq17})(\ref{eq18})(\ref{eq19}) into (\ref{eq20}), the expectation of average waiting time $E[T_{q_1}]$ in the high priority queue is:
\begin{equation}\label{eq21}
\begin{split}
E[T_{q_1}] = & \frac{1}{\lambda_1} {\cdot} {\sum_{n=0}^{n_1}} {\frac{\lambda_1^{n}}{\prod_{i=1}^{n}} [\mu_1+(i-1)\gamma]}    \\
& {\cdot} \frac{n}{1+\sum_{n=1}^{n_1} \prod_{i=1}^{n} \frac{\lambda_1}{\mu_1+(i-1)\gamma}}
\end{split}
\end{equation}Hence, the average reneging probability $P_{reneging}$ for high priority packet is:
\begin{equation}\label{eq22}
P_{reneging}=P_{overflow}+P_{out}
\end{equation}Where, $P_{overflow}$ is the class-1 packet overflow probability in the high priority queue and is defined as:
\begin{equation}\label{eq23}
P_{overflow}= P_{n_1+1}= \frac {{\lambda_1}^{n_1+1}} {\prod_{i=1}^{n_1+1}[\mu_1+(i-1)\gamma]} P_{0}
\end{equation}

\subsection{Low Priority Queue Analysis}
Different from class-1 packets' analysis, class-2 packets are closely in relate with the class-1 queue. Only when the class-1 queue is empty can the lower priority packets be served. Due to the hardness to derive the average waiting time in class-2 queue using the generating function method, we adopt the methods used in \cite{brandt2004two} to give an approximation analysis on the queueing performance of the low priority queue.

Let $\rho_i=\frac{\lambda_i} {\mu_i} (i=1,2)$ denote the submitted load of class-$i$ packets at the cognitive transmission node $SU_{TX}$ in the system and let $\omega$ denote the maximal waiting time of class-1 packets. The expectation to the average number of class-2 packets in the queue at time $t$, $E[n_2(t)] $ can be expressed as \cite{brandt2004two}:
\begin{equation}\label{eq26}
\begin{split}
E[n_2(t)]= & \frac {\mu_2 \rho_2 ( \rho_1-(1-\rho_1)((1+\rho_1)\mu_1 \omega +3)d-d^2)}{\mu_1 (1-\rho_1)((1-\rho_1)-\rho_2(1-d))(1-d)} \\
& + {\frac {\rho_2 (1-d) }{(1-\rho_1)-\rho_2(1-d)}}
\end{split}
\end{equation}
where $d=\rho_1^{2} e^{\frac{\rho_1-1}{\mu_1 \omega}} $. Therefore, the expected average waiting time $E[T_{q_2}] $ can be expressed as \cite{brandt2004two}:
\begin{equation}\label{eq27}
E[T_{q_2}]=\frac{E[n_2(t)]}{\lambda_2}-\frac{1}{\mu_2 p(0)}
\end{equation}
where $p(0) $ is given by (\ref{eq17}).

\section{Simulation Results}\label{sec:simulation}
In order to guarantee the QoS performance of low priority packets when serving the high priority packets in the underlay CRNs, the system parameters should be carefully adjusted. Through simulations, the appropriate parameters could be obtained to guarantee the best system performance. We simulate and analyze the performance of the queueing system using Matlab 7.1.

\begin{figure}[h]\centering
\includegraphics[width=3.0in]{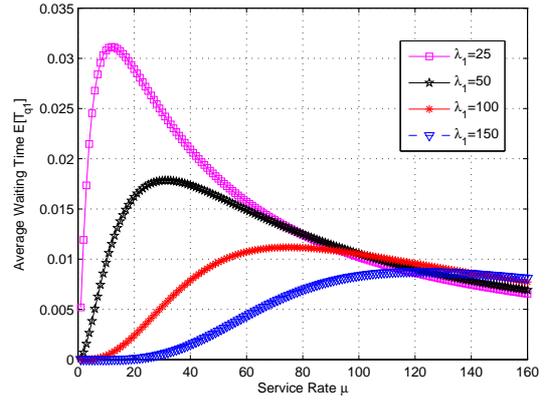}
\caption{Average Waiting Time Vs. Service Rate of High Priority Queue}\label{Fig4}
\end{figure}

The first experiment observes the relationship between the average waiting time in class-1 queue and the service rate of $SU_{TX} $. The interference power is set as $Q=0dB $, then the expectation of average transmission time according to (\ref{eq9}) is about $E[T]=6.2\times10^{-3} $s. Thus, we set the maximum service rate for high priority packet as 160 since $\mu=1/E[T] \approx161.29 $ according to (\ref{eq3}) we set system bandwidth $B=1$MHz, $\gamma=100 $ and $n_1=100 $ as default setting. As we can see from Fig. \ref{Fig4}, the average waiting time increases first and starts to decrease when the service rate $\mu $ increases from 0 to 160 and converges to about 0.008. We also notice that when packet arrival rate $\lambda_1 $ decreases from 150 to 25, the average waiting time increases quickly when service rate falls into the range $[0,60] $. This is because when the packet arrival rate $\lambda_1 $ drops, the number of packets waiting in the priority queue will also decrease. Therefore the reneging packets due to waiting time-out may be lower and finally the actual number of packets waiting to be served in queue may be relatively higher. With the growing number of service rate, the average waiting time of class-1 packets converges to about 0.008 when the service rate approximates 160 which is close to the reciprocal of $E[T]$. This should be explained by that $E[T]$ is a smoothed value of $T$.

\begin{figure}[h]\centering
\includegraphics[width=3.0in]{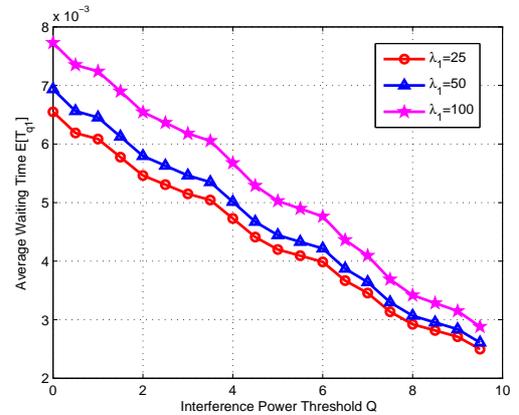}
\caption{Average Waiting Time Vs. Interference Power Threshold}\label{Fig5}
\end{figure}

Next, we observe the impact of the interference power threshold on the average waiting time of high priority queue. The simulation changes the interference power threshold $Q $ from 0dB to 10dB. As shown in Fig. \ref{Fig5}, it is obvious that the average waiting time of high priority queue $E[T_{q_1}] $ is decreasing when the interference power threshold $Q $ becomes larger in the system. That's because when the transmission rate becomes larger at $SU_{TX} $, it will cost much less time to finish the wireless transmission. Fig. \ref{Fig5} also shows that when the interference power threshold is constant, the average waiting time of high priority queue is proportional to the arrival rate of packets $\lambda_1 $.

\begin{figure}[h]\centering
\includegraphics[width=3.0in]{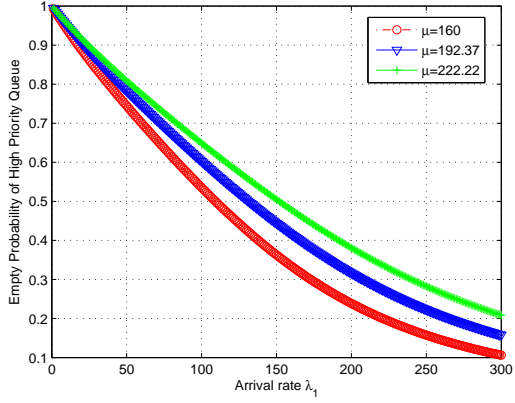}
\caption{Empty probability of high priority queue Vs. packet arrival rate $\lambda_1$}\label{Fig6}
\end{figure}

The third experiment observes the impact of the high priority packet arrival rate $\lambda_1$ on the empty probability of high priority queue. It can be concluded from Fig. \ref{Fig6} that the empty probability of high priority queue is inversely proportional to packet arrival rate $\lambda_1 $. It is easy to understand that the more packets crowding into the high priority queue, the less probability the queue will be empty. Meanwhile, when the packet arrival rate $\lambda_1 $ is constant, the higher the service rate is, the higher chance the empty queue will be.

\begin{figure}[h]\centering
\includegraphics[width=3.0in]{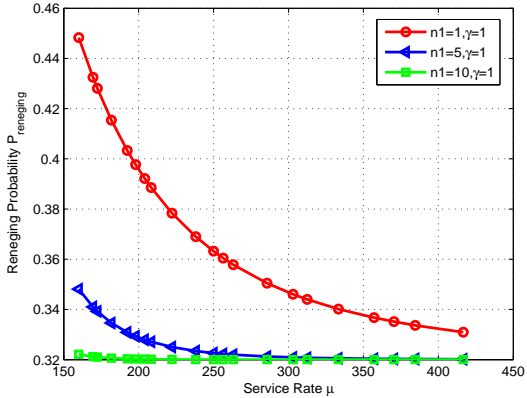}
\caption{Reneging Probability of High Priority Packets Vs. Service Rate $\mu$}\label{Fig7}
\end{figure}

The fourth experiment observes the reneging probability of high priority packets to the service rate of the $SU_{TX} $. It can be observed from Fig. \ref{Fig7} that with the growing number of service rate $\mu $, the probability of reneging $P_{reneging} $ decreases accordingly. This is because when the $SU_{TX} $ is faster to transmit packets to the destination node, the less waiting time the high priority packets will be. Thus, both the number of reneging packets and packets due to waiting time-out will drop which results in a lower reneging probability. Secondly, when the service rate $\mu $ is fixed, with the growing number of maximum allowed queue length $n_1 $, the reneging probability will decrease due to the increasing ability of the high priority queue to handle more incoming packets.

\begin{figure}[h]\centering
\includegraphics[width=3.0in]{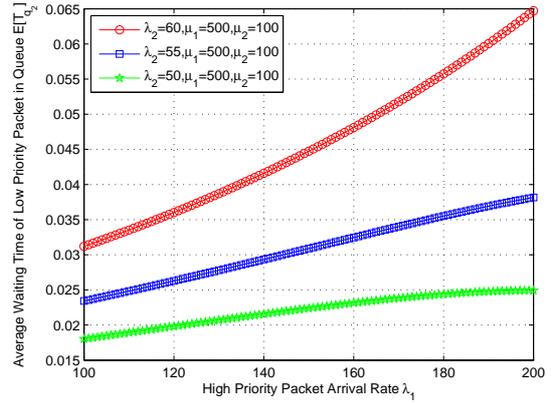}
\caption{Average Waiting Time of Low Priority Packet in Queue Vs. High Priority Packet Arrival Rate}\label{Fig8}
\end{figure}

Finally, we will examine the performance of the low priority queue. As shown in Fig. \ref{Fig8}, the average waiting time of low priority packet in the low priority queue is proportional to the high priority packet arrival rate $\lambda_1 $ when the service rate for class-1 packet $\mu_1=500 $ and the the service rate for class-2 packet $\mu_2=100 $. $\omega$ is set as 0.01. A simple explanation for this is that the average waiting time of class-2 packet grows due to the growing number of high priority packets in class-1 queue. Because packets in class-2 queue can only be served when class-1 queue is empty. We can also observe that when the arrival rate of high priority packet is constant, the average waiting time of low priority packet in the queue is longer when the arrival rate of class-2 packet is larger.

\begin{figure}[h]\centering
\includegraphics[width=3.0in]{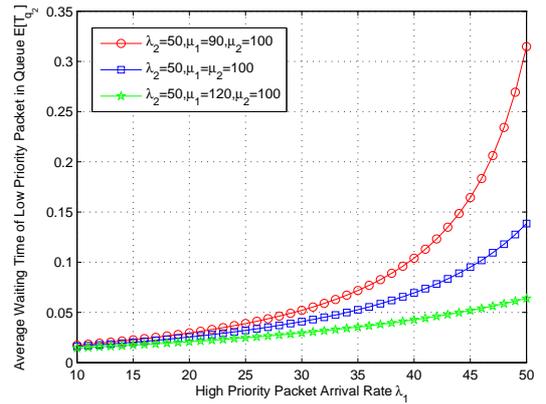}
\caption{Average Waiting Time of Low Priority Packet Vs. High Priority Packet Arrival Rate}\label{Fig9}
\end{figure}

Different from Fig. \ref{Fig8}, when the arrival rate of low priority packets is fixed at $\mu_2=50$, we firstly observe the relationship between the average waiting time of low priority packet in the low priority queue to the arrival rate of high priority queue. The average waiting time of the low priority packets also grows with the growing number of the high priority packets' arrival rate $\lambda_1$. This is depicted in Fig. \ref{Fig9}. Secondly, we observe that, when the service rate of low priority packet $\mu_2$ is fixed at $\mu_2=100$, the average waiting time of low priority packets decreases with the growing service rate of high priority packet $\mu_1$ from $\mu_1=90$ to $\mu_1=120$. Since there are more high priority packets can be transmitted by $SU_{TX}$ when the service rate for class-1 packets is larger, then the number of high priority packets before the class-2 queue will be smaller. This will lead to a shorter average waiting time for the low priority packets.

\section{Conclusion}\label{sec:conclusion}
In this paper, we have proposed a queueing scheme for heterogeneous packets transmissions in underlay cognitive radio networks. In that scheme, emergency or safety related messages possess preemptive higher priority over non-emergency messages under the interference power constraint to primary receive nodes. We model the system behavior as a two dimensional state transition graph and derive the average waiting time, reneging probability of class-1 packets, the expectation of average waiting time in class-2 queue and so on. Through simulations, we demonstrate relationships between queueing system parameters. The analysis of the proposed queueing system in underlay CR network may be applied to cognitive vehicular network system design and other industrial application scenarios. In the future, we will apply the scheme to design time efficient MAC protocols in underlay CR networks. Also, we will add cooperation model to study the influence of cooperative relays in packet level. Since green communication is now becoming more and more popular, in the future, energy consumption model will be constructed to design energy efficient communication schemes in underlay cognitive radio networks.

%
\bibliographystyle{IEEETran}
\bibliography{refs}

\begin{thebibliography}{10}
\providecommand{\url}[1]{#1}
\csname url@samestyle\endcsname
\providecommand{\newblock}{\relax}
\providecommand{\bibinfo}[2]{#2}
\providecommand{\BIBentrySTDinterwordspacing}{\spaceskip=0pt\relax}
\providecommand{\BIBentryALTinterwordstretchfactor}{4}
\providecommand{\BIBentryALTinterwordspacing}{\spaceskip=\fontdimen2\font plus
\BIBentryALTinterwordstretchfactor\fontdimen3\font minus
  \fontdimen4\font\relax}
\providecommand{\BIBforeignlanguage}[2]{{%
\expandafter\ifx\csname l@#1\endcsname\relax
\typeout{** WARNING: IEEEtran.bst: No hyphenation pattern has been}%
\typeout{** loaded for the language `#1'. Using the pattern for}%
\typeout{** the default language instead.}%
\else
\language=\csname l@#1\endcsname
\fi
#2}}
\providecommand{\BIBdecl}{\relax}
\BIBdecl

\bibitem{huang2010beyond}
J.~Huang, G.~Xing, G.~Zhou, and R.~Zhou, ``Beyond co-existence: Exploiting wifi
  white space for zigbee performance assurance,'' in \emph{Network Protocols
  (ICNP), 2010 18th IEEE International Conference on}.\hskip 1em plus 0.5em
  minus 0.4em\relax IEEE, 2010, pp. 305--314.

\bibitem{mitola2000cognitive}
J.~Mitola, ``Cognitive radio: an integrated agent architecture for software
  defined radio,'' Ph.D. dissertation, Royal Institute of Technology (KTH),
  2000.

\bibitem{song2012distributed}
Y.~Song and J.~Xie, ``A distributed broadcast protocol in multi-hop cognitive
  radio ad hoc networks without a common control channel,'' in \emph{INFOCOM,
  2012 Proceedings IEEE}.\hskip 1em plus 0.5em minus 0.4em\relax IEEE, 2012,
  pp. 2273--2281.

\bibitem{ozger2013event}
M.~Ozger and O.~B. Akan, ``Event-driven spectrum-aware clustering in cognitive
  radio sensor networks,'' in \emph{INFOCOM, 2013 Proceedings IEEE}.\hskip 1em
  plus 0.5em minus 0.4em\relax IEEE, 2013, pp. 1483--1491.

\bibitem{han2014throughput}
Y.~Han, E.~Ekici, H.~Kremo, and O.~Altintas, ``Throughput-efficient channel
  allocation in multi-channel cognitive vehicular networks,'' in \emph{INFOCOM,
  2014 Proceedings IEEE}.\hskip 1em plus 0.5em minus 0.4em\relax IEEE, 2014,
  pp. 2724--2732.

\bibitem{srinivasa2007cognitive}
S.~Srinivasa and S.~A. Jafar, ``Cognitive radios for dynamic spectrum
  access-the throughput potential of cognitive radio: A theoretical
  perspective,'' \emph{Communications Magazine, IEEE}, vol.~45, no.~5, pp.
  73--79, 2007.

\bibitem{cheng2011exploiting}
S.-M. Cheng, S.-Y. Lien, F.-S. Chu, and K.-C. Chen, ``On exploiting cognitive
  radio to mitigate interference in macro/femto heterogeneous networks,''
  \emph{Wireless Communications, IEEE}, vol.~18, no.~3, pp. 40--47, 2011.

\bibitem{ghandour2013improving}
A.~J. Ghandour, K.~Fawaz, H.~Artail, M.~Di~Felice, and L.~Bononi, ``Improving
  vehicular safety message delivery through the implementation of a cognitive
  vehicular network,'' \emph{Ad Hoc Networks}, vol.~11, no.~8, pp. 2408--2422,
  2013.

\bibitem{rashid2009opportunistic}
M.~M. Rashid, M.~Hossain, E.~Hossain, and V.~K. Bhargava, ``Opportunistic
  spectrum scheduling for multiuser cognitive radio: a queueing analysis,''
  \emph{Wireless Communications, IEEE Transactions on}, vol.~8, no.~10, pp.
  5259--5269, 2009.

\bibitem{chu2013performance}
T.~M.~C. Chu, H.~Phan, and H.-J. Zepernick, ``On the performance of underlay
  cognitive radio networks using m/g/1/k queueing model,'' \emph{Communications
  Letters, IEEE}, vol.~17, no.~5, pp. 876--879, 2013.

\bibitem{tran2011queuing}
H.~Tran, T.~Q. Duong, and H.~Zepernick, ``Queuing analysis for cognitive radio
  networks under peak interference power constraint,'' in \emph{Wireless and
  Pervasive Computing (ISWPC), 2011 6th International Symposium on}.\hskip 1em
  plus 0.5em minus 0.4em\relax IEEE, 2011, pp. 1--5.

\bibitem{suliman2009queueing}
I.~Suliman and J.~Lehtomaki, ``Queueing analysis of opportunistic access in
  cognitive radios,'' in \emph{Cognitive Radio and Advanced Spectrum
  Management, 2009. CogART 2009. Second International Workshop on}.\hskip 1em
  plus 0.5em minus 0.4em\relax IEEE, 2009, pp. 153--157.

\bibitem{khabazian2011modeling}
M.~Khabazian and S.~Aissa, ``Modeling and performance analysis of cooperative
  communications in cognitive radio networks,'' in \emph{Personal Indoor and
  Mobile Radio Communications (PIMRC), 2011 IEEE 22nd International Symposium
  on}.\hskip 1em plus 0.5em minus 0.4em\relax IEEE, 2011, pp. 598--603.

\bibitem{kam2013multicast}
C.~Kam, S.~Kompella, G.~D. Nguyen, J.~E. Wieselthier, and A.~Ephremides,
  ``Multicast throughput stability analysis for cognitive cooperative random
  access,'' in \emph{INFOCOM, 2013 Proceedings IEEE}.\hskip 1em plus 0.5em
  minus 0.4em\relax IEEE, 2013, pp. 170--174.

\bibitem{chang2013spectrum}
W.~S. Chang and W.~M. Jang, ``Spectrum occupancy of cognitive radio networks: a
  queueing analysis using retrial queue,'' \emph{IET Networks}, vol.~3, no.~3,
  pp. 218--227, 2013.

\bibitem{wang2013belief}
Y.~Wang, H.~Li, and L.~Qian, ``Belief propagation based spectrum sensing
  subject to dynamic primary user activities: Phantom of quickest detection,''
  in \emph{Military Communications Conference, MILCOM 2013-2013 IEEE}.\hskip
  1em plus 0.5em minus 0.4em\relax IEEE, 2013, pp. 1193--1200.

\bibitem{garcia2012stochastic}
C.~Garcia-Costa, E.~Egea-Lopez, J.~B. Tomas-Gabarron, J.~Garcia-Haro, and Z.~J.
  Haas, ``A stochastic model for chain collisions of vehicles equipped with
  vehicular communications,'' \emph{Intelligent Transportation Systems, IEEE
  Transactions on}, vol.~13, no.~2, pp. 503--518, 2012.

\bibitem{yu2007self}
F.~Yu and S.~Biswas, ``Self-configuring tdma protocols for enhancing vehicle
  safety with dsrc based vehicle-to-vehicle communications,'' \emph{Selected
  Areas in Communications, IEEE Journal on}, vol.~25, no.~8, pp. 1526--1537,
  2007.

\bibitem{guo2013r}
W.~Guo, L.~Huang, L.~Chen, H.~Xu, and C.~Miao, ``R-mac: Risk-aware dynamic mac
  protocol for vehicular cooperative collision avoidance system,''
  \emph{International Journal of Distributed Sensor Networks}, vol. 2013, 2013.

\bibitem{ross2014introduction}
S.~M. Ross, \emph{Introduction to probability models}.\hskip 1em plus 0.5em
  minus 0.4em\relax Academic press, 2014.

\bibitem{klennrock1975queueing}
L.~Klennrock, ``Queueing systems volume 1: theory,'' \emph{New York}, 1975.

\bibitem{gong2006queue}
Q.~Gong and R.~Batta, ``A queue-length cutoff model for a preemptive
  two-priority m/m/1 system,'' \emph{SIAM Journal on Applied Mathematics},
  vol.~67, no.~1, pp. 99--115, 2006.

\bibitem{khan2012delay}
F.~A. Khan, K.~Tourki, M.~Alouini, and K.~A. Qaraqe, ``Delay analysis of a
  point-to-multipoint spectrum sharing network with csi based power
  allocation,'' in \emph{Dynamic Spectrum Access Networks (DYSPAN), 2012 IEEE
  International Symposium on}.\hskip 1em plus 0.5em minus 0.4em\relax IEEE,
  2012, pp. 235--241.

\bibitem{tran2013cognitive}
H.~Tran, H.-J. Zepernick, and H.~Phan, ``Cognitive proactive and reactive df
  relaying schemes under joint outage and peak transmit power constraints,''
  \emph{Communications Letters, IEEE}, vol.~17, no.~8, pp. 1548--1551, 2013.

\bibitem{tran2012delay}
H.~Tran, T.~Q. Duong, and H.-J. Zepernick, ``Delay performance of cognitive
  radio networks for point-to-point and point-to-multipoint communications,''
  \emph{EURASIP Journal on Wireless Communications and Networking}, vol. 2012,
  no.~1, pp. 1--15, 2012.

\bibitem{brandt2004two}
A.~Brandt and M.~Brandt, ``On the two-class m/m/1 system under preemptive
  resume and impatience of the prioritized customers,'' \emph{Queueing
  Systems}, vol.~47, no. 1-2, pp. 147--168, 2004.

\end{thebibliography}

\end{document}